\documentclass[conference]{IEEEtran}
\IEEEoverridecommandlockouts

\usepackage[T1]{fontenc}
\usepackage{cite}
\usepackage{amsmath,amssymb,amsfonts}
\usepackage{amsthm}
\usepackage{graphicx}
\usepackage{textcomp}
\usepackage{xcolor}
\usepackage[caption=false,font=footnotesize]{subfig}
\usepackage{capt-of}
\usepackage[ruled,vlined,linesnumbered]{algorithm2e}
\SetAlCapFnt{\footnotesize}
\SetAlCapNameFnt{\footnotesize}
\usepackage{multirow}
\usepackage{array}
\usepackage{tablefootnote}
\usepackage{pifont}
\usepackage{threeparttable}
\usepackage{tikz}
\usepackage[hidelinks,pdfusetitle]{hyperref}

\newcommand{\EMAMultiMin}{1.20}
\newcommand{\EMAMultiMax}{28.18}
\newcommand{\EMAMultiMean}{5.69}

\newcommand{\EMAUpdateRetention}{98.03}

\hypersetup{
  pdfauthor={Mocheng Li, Xiao Yan, Baotong Lu, James Cheng, Chenhao Ma}
}

\newtheorem{theorem}{Theorem}[section]
\newtheorem{lemma}[theorem]{Lemma}
\theoremstyle{definition}
\newtheorem{definition}[theorem]{Definition}

\newcolumntype{Z}[1]{>{\centering\arraybackslash}p{#1}}

\newcommand{\graycircled}[1]{%
\tikz[baseline=(char.base)]{
\node[shape=circle, draw=gray, fill=gray!20, inner sep=1pt]
(char) {\small\bfseries #1};}}

\title{EMA: Approximate Nearest Neighbor Search with General Attribute Filtering and Dynamic Updates}

\author{
\IEEEauthorblockN{\makebox[0.3\textwidth]{\href{https://orcid.org/0000-0002-5524-1042}{Mocheng Li}}}
\IEEEauthorblockA{\textit{The Chinese University of}\\
\textit{Hong Kong, Shenzhen}\\
Shenzhen, China\\
mochengli1@link.cuhk.edu.cn}
\and
\IEEEauthorblockN{\makebox[0.3\textwidth]{\href{https://orcid.org/0000-0002-2122-915X}{Xiao Yan}}}
\IEEEauthorblockA{\textit{Institute for Math \& AI}\\
\textit{Wuhan University}\\
Wuhan, China\\
yanxiaosunny@whu.edu.cn}
\and
\IEEEauthorblockN{\makebox[0.3\textwidth]{\href{https://orcid.org/0000-0002-0230-1048}{Baotong Lu}}}
\IEEEauthorblockA{\textit{Microsoft Research}\\
\strut\\
Beijing, China\\
baotonglu@microsoft.com}
\and[\hfill\mbox{}\linebreak\mbox{}\hfill]
\IEEEauthorblockN{\makebox[0.3\textwidth]{\href{https://orcid.org/0000-0001-6313-6288}{James Cheng}}}
\IEEEauthorblockA{\textit{The Chinese University of}\\
\textit{Hong Kong}\\
Hong Kong, China\\
jcheng@cse.cuhk.edu.hk}
\and
\IEEEauthorblockN{\makebox[0.3\textwidth]{\href{https://orcid.org/0000-0002-3243-8512}{Chenhao Ma}\textsuperscript{*}}%
\thanks{\textsuperscript{*}Corresponding author: Chenhao Ma.}}
\IEEEauthorblockA{\textit{The Chinese University of}\\
\textit{Hong Kong, Shenzhen}\\
Shenzhen, China\\
machenhao@cuhk.edu.cn}
}

\begin{document}
\maketitle

\begin{abstract}
Filtering approximate nearest neighbor search (Filtering ANNS) retrieves approximate nearest neighbors subject to attribute constraints in applications such as recommendation, retrieval-augmented generation (RAG), and LLM agent memory.
However, predicate-specialized methods target either range or label filtering and can incur substantial index construction time, whereas predicate-agnostic methods support arbitrary predicates but struggle to sustain throughput and recall across a wide range of selectivities.
In this paper, we propose \textbf{EMA}, a Filtering ANNS algorithm that supports multi-predicate queries over mixed numerical and categorical attributes as well as efficient dynamic updates.
EMA attaches compact summaries, called \emph{Markers}, to graph edges to record the attributes of candidates removed by geometric pruning.
During query processing, EMA performs Marker-guided joint search with bounded edge recovery, enabling efficient filtering while preserving graph navigability.
On multi-predicate workloads, EMA achieves \textbf{\EMAMultiMin\texttimes{}--\EMAMultiMax\texttimes{} the query throughput} of the best competing baseline at comparable recall.
In our complete-record replacement experiment, EMA retains approximately 98\% of its initial query throughput without rebuilding the graph.
\end{abstract}


\section{Introduction}

\begin{figure}[!t]
    \centering
    \subfloat[Classical RNG pruning.\label{subfig:rngprune}]{%
    \begin{minipage}[b]{0.4\linewidth}
        \centering
        \includegraphics[width=0.85\linewidth]{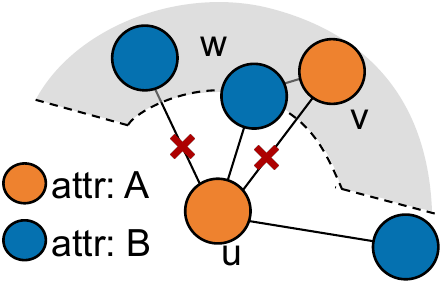}
    \end{minipage}}%
    \hfill
    \subfloat[Joint filtering.\label{subfig:rngjoint}]{%
    \begin{minipage}[b]{0.5\linewidth}
        \centering
        \includegraphics[width=0.85\linewidth]{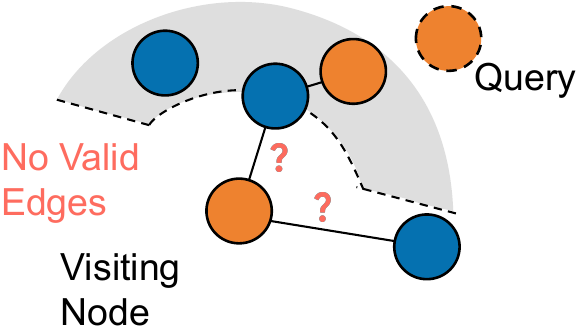}
    \end{minipage}}%
    \par
    \subfloat[ACORN pruning.\label{subfig:acornprune}]{%
    \begin{minipage}[b]{0.4\linewidth}
        \centering
        \includegraphics[width=0.6\linewidth]{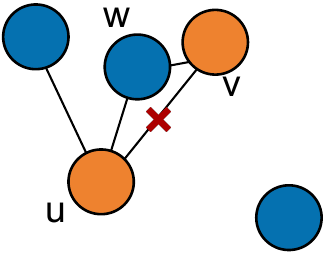}
    \end{minipage}}%
    \hfill
    \subfloat[Filtered DiskANN pruning.\label{subfig:diskannprune}]{%
    \begin{minipage}[b]{0.5\linewidth}
        \centering
        \includegraphics[width=0.6\linewidth]{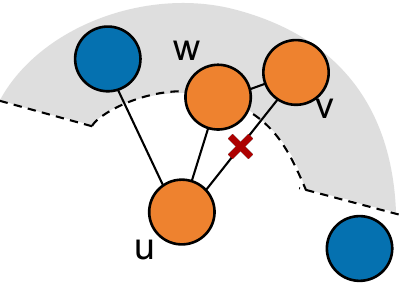}
    \end{minipage}}%
    \caption{Graph pruning and traversal under attribute filtering.}
    \label{fig:rng}
\end{figure}

Embedding-based retrieval is widely used in retrieval-augmented generation (RAG)~\cite{edge2024local, zhou2025depth, wang2025archrag}, recommendation systems~\cite{meng2020pmd, okura2017embedding, paterek2007improving}, semantic search~\cite{patel2024lotus}, and LLM agent memory~\cite{packer2023memgpt, park2023generative, chhikara2025mem0}.
Queries often combine numerical ranges, such as a document's effective date, with categorical constraints on tags or other metadata.
An LLM agent, for example, may filter memories by user ID and time window while continuously adding, revising, and deleting memory records~\cite{chhikara2025mem0}.
Filtering approximate nearest neighbor search (Filtering ANNS) retrieves approximate nearest neighbors within the subset of records satisfying such predicates.
Supporting these mixed predicates requires graph traversal to account for attribute constraints while efficiently locating geometrically close results.

Graph-based indexes, such as HNSW~\cite{malkov2018efficient} and DiskANN~\cite{jayaram2019diskann}, provide an effective foundation for ANN search.
They retain a small set of geometrically diverse neighbors and navigate the resulting sparse graph toward the query.
In Figure~\ref{subfig:rngprune}, RNG-style pruning removes the edge from $u$ to $v$ because retained neighbor $w$ geometrically dominates $v$.
With predicate checks on directly reached neighbors, the search queue admits matching candidates.
If $u$ satisfies the predicate but $w$ does not, search cannot continue through $w$; Figure~\ref{subfig:rngjoint} illustrates how this restriction blocks an expansion direction.
Post-filtering approaches, including VBase~\cite{zhang2023vbase}, permit non-matching intermediates, retaining routing flexibility at the cost of exploring non-matching branches.
Under restrictive predicates, this exploration can consume substantial work before enough valid results are found.

Predicate-specialized indexes, such as iRangeGraph~\cite{xu2024irangegraph} for numerical ranges and Filtered DiskANN~\cite{gollapudi2023filtered} for labels, incorporate attribute constraints into index construction.
Their graph structures and pruning rules are tailored to numerical ranges or label membership.
For example, Filtered DiskANN conditions geometric pruning on label coverage, retaining an edge when the replacement neighbor does not cover the required labels (Figure~\ref{subfig:diskannprune}).
Predicate-agnostic Filtering ANNS methods, including ACORN~\cite{patel2024acorn} and NaviX~\cite{sehgal2025navix}, support mixed constraints through joint filtering and traversal.
ACORN's pruning considers alternative two-hop paths when removing candidate edges (Figure~\ref{subfig:acornprune}).
During search, these methods use additional neighbor exploration to discover matching candidates.
Such exploration accommodates changing predicates, but incurs graph and attribute accesses to determine useful expansion directions.

These considerations motivate three related challenges for a filtered graph index.

\graycircled{1} \textbf{Selective Expansion with Routing Flexibility.}
How can search avoid expanding large numbers of non-matching records while retaining useful non-matching intermediates?
A non-matching neighbor may still provide a useful route to matching candidates beyond it, while its own attributes offer little guidance about whether such candidates exist.
Search therefore needs inexpensive evidence to identify promising expansion directions.

\graycircled{2} \textbf{Pruning-derived Attribute Evidence.}
How can attributes of candidates removed by geometric pruning remain available for query-time guidance without additional neighborhood exploration to collect them?
RNG-style neighbor selection can remove $(u,v)$ when $v$ is dominated by a retained neighbor $w$ under its distance-based pruning rule.
Yet $w$ need not satisfy a query predicate $R$ that $v$ satisfies.
Once $(u,v)$ is removed, checking $w$'s attributes alone cannot reveal $v$'s compatibility with $R$.
The pruning relationship provides a natural association between the pruned candidate $v$ and the retained expansion direction $(u,w)$.
The problem is to encode this pruning-derived association as compact edge-level evidence, making the attributes of pruned candidates available during traversal without restoring them as neighbors of the source node.

\graycircled{3} \textbf{Maintaining Graph and Filtering Evidence Dynamically.}
Once attribute evidence is distributed across retained edges, how can the graph and its filtering guidance be maintained under insertions, deletions, and complete-record replacements without rebuilding the entire graph?
Deletion can affect both incident edges and attribute contributions previously transferred to other edges.
Local maintenance must account for these dispersed contributions as well as changes to the routing structure.

We propose Edge Marker (EMA)\footnote{The source code, data, and/or other artifacts have been made available at \url{https://github.com/lmccccc/EMA}.}, a graph-based Filtering ANNS framework that attaches compact attribute summaries called \emph{Markers} to retained edges.

To enable selective expansion, EMA checks edge Markers before dereferencing neighboring records, using predicate-compatible evidence to guide expansion through non-matching intermediates.
Bounded edge recovery supplies additional expansion opportunities under restrictive predicates.
Accessed records undergo exact predicate verification.

To preserve pruning-derived evidence, EMA records attribute information from geometrically pruned candidates in the Markers of their corresponding retained edges during construction.
At query time, these Markers expose the transferred evidence through the retained graph without restoring the pruned candidates as neighbors of the source node.
Codebooks encode numerical regions and categorical labels into per-attribute bit summaries, allowing combinations of \texttt{AND}/\texttt{OR} predicates to use the same graph.

For dynamic maintenance, EMA jointly maintains graph edges and their Markers under insertion, deletion, and complete-record replacement.
Local graph repair, Marker cleanup, and reconstruction update the routing structure and its filtering evidence without rebuilding the entire graph.

Our main contributions are summarized as follows:
\begin{itemize}
    \item We propose EMA, an edge-centric Filtering ANNS framework that preserves pruning-derived attribute evidence to guide traversal for mixed numerical and categorical predicates on a shared graph.

    \item We develop local graph and Marker maintenance for insertion, deletion, and complete-record replacement. We analyze construction, storage, and local deletion costs, and derive source-specific Marker false-positive bounds under explicit assumptions.

    \item We evaluate multi-predicate and negatively correlated queries, dynamic maintenance, and resource overhead, demonstrating EMA's query benefits, performance under continued updates, and workload-dependent design trade-offs.
\end{itemize}

\section{Preliminaries}
\label{sec:preliminaries}

Here, we first define the ANN and Filtering ANNS problems and then review graph-based ANN indexes and the RNG pruning principle that underlies their connectivity.

\subsection{Problem Definition}

Let $\mathcal{D} = (\mathcal{V}, \mathcal{A})$ denote a dataset, where $\mathcal{V}$ is a set of $d$-dimensional vectors and $\mathcal{A}$ is the associated attribute set.
Each vector $v_i \in \mathcal{V}$ is associated with an attribute record $A_i \in \mathcal{A}$.
Given a query $q = (v, R)$ consisting of a query vector $v$ and an attribute predicate $R$, the goal of Filtering ANNS is to retrieve vectors that are both close to $v$ and satisfy $R$.

The approximate nearest neighbor (ANN) search aims to retrieve $k$ vectors $\{v_{i}\}$ from $\mathcal{V}$ whose distances to $v$ approximately minimize $\mathrm{dis}(v, v_i)$.
Filtering ANNS further restricts the search space to the filtered subset
\[
\mathcal{V}_R = \{ v_i \in \mathcal{V} \mid R(A_i) = \mathrm{true} \},
\]
and returns approximate nearest neighbors within $\mathcal{V}_R$.

\begin{table}[t]
  \caption{Notations in our paper.}
  \label{tab:notation}
  \begin{tabular}{l|l}
    \hline
    \textbf{Notations} & \textbf{Descriptions}  \\
    \hline
    $\mathcal{D}=(\mathcal{V},\mathcal{A})$ & Dataset with vector set $\mathcal{V}$ and attribute set $\mathcal{A}$. \\
    \hline
    $q=(v,R)$ & Query vector and predicate(s). \\
    \hline
    $m$ & Cardinality of attribute set $\mathcal{A}$. \\
    \hline
    $\mathcal{A}^{(j)}$ & The set of $j$-th attribute in $\mathcal{A}$, $0<j \le m$. \\
    \hline
    $A_i^{(j)}$ & The $j$-th attribute of item $i$, $0<j \le m$. \\
    \hline
    $dis(v_i,v_j)$ & Distance between vector $v_i$ and $v_j$. \\
    \hline
    $d$ & Dimension per vector in $\mathcal{V}$. \\
    \hline
    $n$ & Size of dataset $\mathcal{D}$. \\
    \hline
    $M$ & Bottom-layer maximum degree ($2M_{\mathrm{HNSW}}$). \\
    \hline
    $efc$ & Candidate set size in index construction. \\
    \hline
    $s$ & Codebook buckets (Marker bits) per attribute. \\
    \hline
    $d_{\min}$ & Edge-recovery threshold. \\
    \hline
  
\end{tabular}
\end{table}

Table~\ref{tab:notation} summarizes the frequent notations used throughout this paper.

\subsection{Graph-based ANN and RNG Pruning}

Graph-based approximate nearest neighbor (ANN) methods organize data points into a proximity graph and perform greedy graph traversal during search, achieving strong performance in high-dimensional spaces.
Representative systems include HNSW~\cite{malkov2018efficient}, DiskANN~\cite{jayaram2019diskann}, and FANNG~\cite{harwood2016fanng}.
Despite differences in index construction and search strategies, these methods share a common structural principle: maintaining a sparse yet navigable graph through carefully designed neighbor pruning rules.

A widely adopted pruning strategy is inspired by the Relative Neighborhood Graph (RNG)~\cite{lankford1968regionalization,toussaint1980relative}, which promotes structural diversity by removing neighbors that are dominated by closer ones.
Formally, given a source node $u$, a neighbor $v$ is dominated by another neighbor $w$ if
\[
\mathrm{dis}(u,v) > \mathrm{dis}(u,w)
\quad \text{and} \quad
\mathrm{dis}(w,v) < \mathrm{dis}(u,v).
\]
Geometrically, as illustrated in Figure~\ref{subfig:rngprune}, fixing a source node $u$ and one of its neighbors $w$, the neighbors $v$ that can be dominated by $w$ lie within a truncated half-space determined by the perpendicular bisector of $(u,w)$.
This geometric property preserves graph sparsity while maintaining diversity among retained neighbors. 
In practice, graph-based ANN indexes commonly adopt \emph{RNG-style} pruning rules that relax this classical condition for efficiency, while preserving its diversification intuition.

RNG-style pruning bases neighbor selection on geometric proximity.
With predicate checks on directly reached neighbors, geometric pruning can block expansion toward matching records.
In Figure~\ref{subfig:rngjoint}, all retained neighbors of a predicate-satisfying source $u$ violate the predicate, leaving no eligible edge for traversal toward nearby matching records.

Several existing methods attempt to mitigate this issue by relaxing RNG-style pruning.
ACORN~\cite{patel2024acorn} adopts a two-hop pruning strategy, as illustrated in Figure~\ref{subfig:acornprune}, where a candidate neighbor is pruned only if it remains reachable through an alternative two-hop path.

Filtered DiskANN~\cite{gollapudi2023filtered} introduces attribute-aware pruning by removing an edge only when the retained neighbor covers all labels shared by both endpoints.
Specifically, RNG-style pruning is applied only if $u.A \cap v.A \subseteq w.A$, as shown in Figure~\ref{subfig:diskannprune}.
When label sets differ, this coverage condition can prevent the removal of geometrically dominated candidates, affecting neighbor selection under a bounded degree budget.

In summary, \emph{RNG-style pruning} provides geometric sparsification, but filtering can restrict traversal through the retained neighbors.
This motivates retaining attribute evidence alongside geometric pruning to guide selective traversal.

\section{EMA}

This section presents the design of EMA, a graph-based index for filtering-aware approximate nearest neighbor search.
EMA extends RNG-style pruning with a compact \emph{Marker} that encodes predicate-relevant neighborhood information on each edge, enabling efficient joint navigation under attribute filtering.



\begin{figure*}[t]
    \centering
    \includegraphics[width=0.85\linewidth]{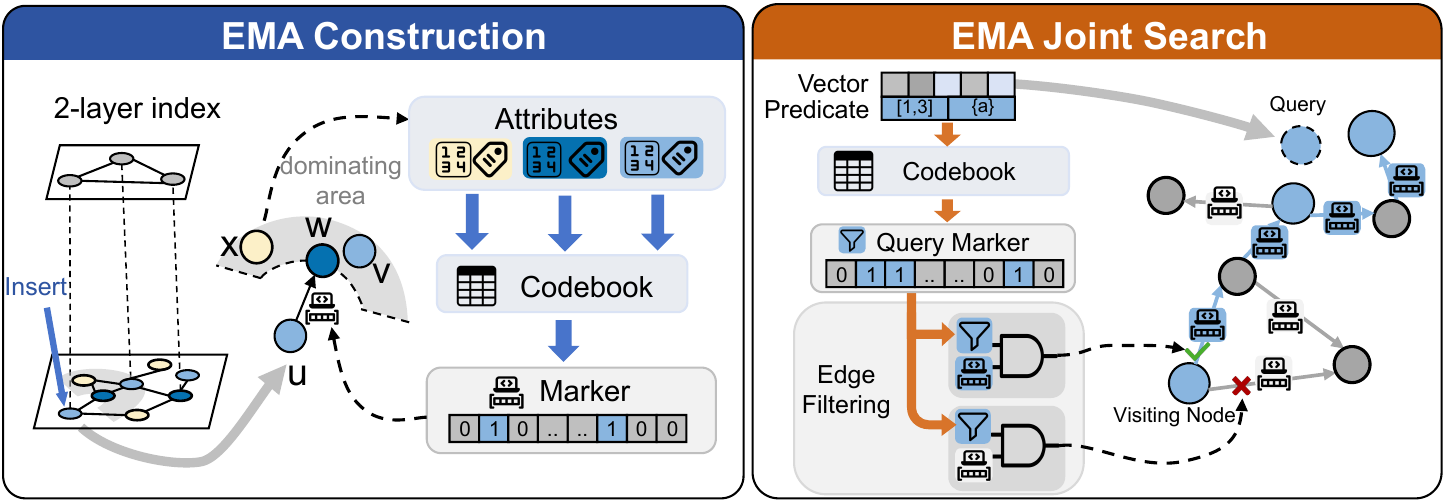}
    \caption{EMA workflow.}
    
    \label{fig:workflow}
\end{figure*}

As illustrated in Figure~\ref{fig:workflow}, EMA organizes HNSW-style search and pruning in two graph layers: an upper navigation layer (layer~1) and a bottom search layer (layer~0).
Incremental construction uses candidate search and RNG-style neighbor pruning.
Markers reside on bottom-layer edges, aggregating attributes from nearby dominated nodes during pruning.
Markers thus summarize attribute information associated with geometry-aware neighborhoods.
At query time, vector-distance navigation in layer~1 locates entry points for Marker-guided joint search in layer~0.
Edges are selectively traversed based on Marker–predicate compatibility, dynamically inducing a predicate-specific subgraph for efficient filtering-aware ANN search.

\subsection{Core Abstractions: Marker and Codebook}

To support efficient joint search under general attribute filtering, EMA introduces two core abstractions: the \emph{Marker} and the \emph{Codebook}.
Together, they provide a compact and conservative edge-centric, hop-agnostic representation of predicate-relevant attribute information associated with local graph neighborhoods, enabling predicate prechecking over the encoded attribute evidence during search.
We next formalize their definitions and encoding mechanisms.

\begin{definition}[Marker]
A \emph{Marker} encodes the presence of attribute values associated with a subset of vectors $S \subseteq \mathcal{V}$ with respect to the global attribute set.

Let $A = \{A^{(1)}, \dots, A^{(m)}\}$ denote the set of attributes of a vector, 
where $A^{(j)}$ is a set of values for categorical attributes and a singleton for numerical attributes.
Given a subset of attribute values
$\{A_i^{(j)} \mid v_i \in S\}$,
a fixed-length bit array
$\mathrm{Marker}_j \in \{0,1\}^{s}$ is constructed for attribute $A^{(j)}$,
where each bit position corresponds to a discretized value region.
A bit is set to $1$ if there exists at least one vector in $S$ whose attribute value falls into the corresponding region.

The overall Marker for subset $S$ is obtained by concatenating the per-attribute Markers:
\par
\begingroup
\small
\[
S.\mathrm{Marker} =
\bigl[
\mathrm{Marker}_1 \;\|\;
\mathrm{Marker}_2 \;\|\; \cdots \;\|\;
\mathrm{Marker}_{m}
\bigr] \in \{0,1\}^{s \cdot m}.
\]
\endgroup
\end{definition}

Due to discretization and aggregation, Markers may admit false positives during predicate checking.
The encoding preserves predicate compatibility for the records summarized by a Marker, supporting conservative checks without introducing extra false negatives for those records.

To construct and interpret Markers, EMA requires a unified representation that maps heterogeneous attribute values into a compact bit-level form.
We therefore introduce a \emph{Codebook} to define how attribute domains are discretized and encoded.

\begin{definition}[Codebook]
Given a dataset $\mathcal{D} = (\mathcal{V}, \mathcal{A})$, a \emph{Codebook}
$\mathcal{C} = \{C_j\}_{j=1}^{m}$ is a collection of attribute-specific discretization functions, where each
\[
C_j : \mathcal{X}_j \rightarrow \{1,\dots,s\}
\]
maps values from the domain $\mathcal{X}_j$ of attribute $A^{(j)}$ to one of $s$ discrete buckets.
For categorical attributes, $\mathcal{X}_j$ is the registered label set; it can grow within the configured capacity without changing existing assignments.
\end{definition}



The Codebook provides a unified and fixed-size discretization scheme for heterogeneous attributes, enabling different attribute types to be encoded into a unified bit-level representation.

\begin{figure}[t]
    \centering
    \includegraphics[width=\linewidth]{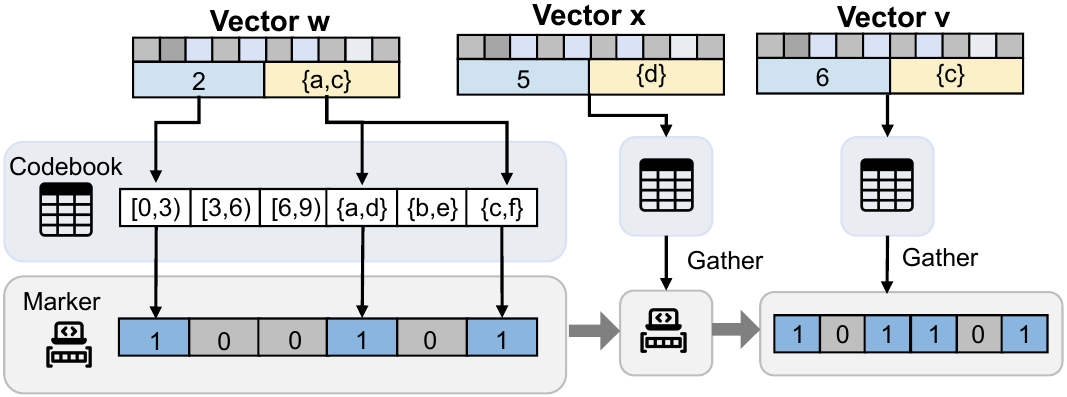}
    \caption{Illustration of Codebook-based Marker encoding and gathering.}
    \label{fig:marker}
\end{figure}

\subsection{EMA Construction}
\label{sec:ema-construction}

EMA incrementally constructs its two-layer graph using HNSW-style candidate search and degree-bounded neighbor pruning.
Bottom-layer pruning additionally aggregates attribute evidence into edge Markers.
We first describe Codebook construction, which precedes graph insertion.

\begin{algorithm}[ht]
\caption{{\tt Codebook Generation}}
\label{alg:Codebook}
\KwIn{Attribute columns $\{\{A_i^{(j)}\}_i\}_{j=1}^{m}$, number of buckets $s$}
\KwOut{Codebook $\mathcal{C}=\{C_j\}_{j=1}^{m}$}
\small
\For{$j \gets 1$ \KwTo $m$}{
    \eIf{attribute $j$ is numerical}{
        sort values in $\{A_i^{(j)}\}_i$\;
        partition into $s$ contiguous buckets\;
        define $C_j$ using bucket lower bounds\;
    }{
        compute category frequencies in $\{A_i^{(j)}\}_i$\;
        sort observed categories by frequency\;
        assign each to a bucket with the fewest assigned labels\;
        define $C_j$ as a category-to-bucket mapping\;
    }
}
\end{algorithm}



\textbf{Codebook generation.}
Before index construction, EMA initializes a Codebook for each attribute.
For attribute $j$, the Codebook is constructed from the empirical distribution of all values
$\{A_i^{(j)} \mid v_i \in \mathcal{V}\}$ in the dataset, yielding a deterministic mapping shared across index construction and query processing.

Algorithm~\ref{alg:Codebook} outlines the procedure.
Numerical attributes are discretized by sorting values and partitioning them into $s$ contiguous frequency-balanced buckets, with bucket boundaries defining the mapping.
Observed categorical labels are processed by frequency and assigned to buckets balancing the number of labels, inducing a category-to-bucket mapping.


\textbf{Attribute encoding.}
With the Codebook defined, we describe how Markers are constructed during index insertion.
Figure~\ref{fig:marker} illustrates Codebook-based Marker encoding and aggregation for both numerical and categorical attributes.

Algorithm~\ref{alg:dtpassing} formalizes the encoding process.
For each attribute, its value (or values, for categorical attributes) is mapped to a bucket index using the Codebook, and the corresponding bit in the Marker is set.

\begin{algorithm}[ht]
\caption{{\tt MEncode}}
\label{alg:dtpassing}
\small
\KwIn{Attribute set $A$, Codebook $\mathcal{C}$}
\KwOut{Marker $Marker$}
\lFor{$j \gets 1$ to $m$}{
    $\text{Marker}_j \gets 0^s$
}
\ForEach{attribute $j = 1,\dots,m$}{
    \ForEach{value $a \in A^{(j)}$}{
        $b \gets C_j(a)$\;
        $Marker_j[b] \gets 1$\;
    }
}
\Return{$\operatorname{Concat}_{j=1}^{m}(\text{Marker}_j)$}\;
\end{algorithm}

\textbf{Marker construction with RNG-style pruning.}
As a core component of graph-based ANN indexes, RNG-style pruning plays a crucial role in maintaining a geometrically sparse, well-connected, and diverse neighborhood structure.
While \texttt{MEncode} enables attribute information to be encoded on individual edges, naive pruning would discard such information together with the pruned edges.
This would eliminate attribute cues associated with the dominated regions and degrade the navigability of the graph under filtering constraints.
To address this issue, EMA propagates the Marker information from pruned edges to the retained edges that dominate them, thereby preserving predicate-relevant routing evidence after graph sparsification.

As illustrated in Figure~\ref{fig:workflow}, when inserting node $u$ into the bottom layer,
multiple candidate nodes (e.g., $x$ and $v$) may fall within the dominating region of a selected neighbor $w$.
During RNG-style neighbor pruning, instead of implicitly losing the attribute information associated with the pruned edges $(u,x)$ and $(u,v)$,
EMA aggregates their Markers into the Marker attached to the surviving edge $(u,w)$.
As a result, the edge $(u,w)$ compactly represents both geometric proximity and the possible presence of attribute values within its dominated region,
thereby preserving predicate-relevant information for subsequent navigation.
The detailed Codebook-based Marker encoding and aggregation mechanism is illustrated in Figure~\ref{fig:marker}.

Importantly, because this propagation follows RNG-style domination defined by local geometric relations (pairwise distances),
the resulting Marker aggregation is hop-agnostic and geometry-aware: attribute evidence is gathered during pruning without multi-hop exploration and is inherently geometry-dependent.

\begin{algorithm}[t]
\caption{{\tt Marker-augmented RNG Pruning}}
\label{alg:newpruning}
\small
\KwIn{Node $u$; distance-sorted $Cand$ with $|Cand|\ge M$; layer budget $M$; Codebook $\mathcal{C}$}
\KwOut{Neighbor list $Nbrs$; Markers on edges incident to $u$.}

$Nbrs \gets \emptyset$\;

\ForEach{$v \in Cand$}{
    $dom \gets \mathrm{false}$\;

    \ForEach{$w \in Nbrs$}{
        \If{$dis(w,v) < dis(u,v)$}{ 
            \uIf{$e_{(u,v)}$ is an old edge}{
                $e_{(u,w)}.\mathrm{Marker} \mathrel{|\!=} e_{(u,v)}.\mathrm{Marker}$\;
            }
            \Else{
                $e_{(u,w)}.\mathrm{Marker} \mathrel{|\!=} {\tt MEncode}(v.A,\mathcal{C})$\;
            }
            $dom \gets \mathrm{true}$\;
            \textbf{break}\;
        }
    }

    \If{$dom = \mathrm{false}$}{
        \lIf{$|Nbrs| = M$}{\textbf{break}}
        $Nbrs \gets Nbrs \cup \{v\}$\;
        $e_{(u,v)}.\mathrm{Marker} \mathrel{|\!=} {\tt MEncode}(v.A,\mathcal{C})$\;
    }
}
\end{algorithm}

\textbf{Pruning algorithm design.}
Algorithm~\ref{alg:newpruning} presents the core Marker-propagation procedure during bottom-layer neighbor pruning.
Complete construction settings used in the experiments are provided in the technical report~\cite{ema_appendix}.
Similar to HNSW, EMA maintains a candidate list $Cand$ obtained by greedy search and prunes it to satisfy an out-degree budget $M$.
We assume $Cand$ is sorted in ascending order of $dis(u,\cdot)$.
If $|Cand|<M$, all candidates are retained without pruning, with target-own initialization for new edge Markers.

\emph{RNG-style domination and Marker propagation.}
For each candidate $v$, we test whether it is dominated by an already selected neighbor $w \in Nbrs$ using an RNG-style condition:
$w$ dominates $v$ if $dis(w,v) < dis(u,v)$.
If $v$ is dominated, the edge $(u,v)$ will be pruned; however, its attribute information should not be lost.
EMA therefore propagates the Marker information to the dominating edge $(u,w)$ by updating $e_{(u,w)}.\mathrm{Marker}$ with a bitwise OR.
When $(u,v)$ is an old edge from earlier point insertions, we reuse its existing Marker; otherwise, we encode $v$'s attributes via \texttt{MEncode} and merge the resulting bit vector.

Nondominated candidates are retained up to the degree budget, and each retained edge includes its target's encoding.
At capacity, dominated candidates may still contribute Markers until the next nondominated candidate terminates pruning.

\textbf{Pruning-derived routing evidence.}
For a retained edge $e=(u,w)$, let $S_e$ denote the records whose encodings have contributed to its Marker during static construction.
Writing $F(x)={\tt MEncode}(x.A,\mathcal{C})$, we have
\[
w\in S_e,\qquad M_e=\bigvee_{x\in S_e}F(x).
\]
The set starts with the target and accumulates contributions from candidates assigned to this edge by pruning.
Transferring an old edge's Marker also carries its earlier contributions, so $S_e$ can include records beyond those directly dominated by the current target.
This conceptual set explains the stored summary; construction maintains the bit vector rather than contributor identities.

Suppose candidate $v$ satisfies a query predicate but is pruned in favor of a closer, non-matching neighbor $w$.
Checking only $w$'s attributes would reject this direction.
After $v$'s encoding is transferred to $u\!\to\!w$, the Marker retains its predicate-compatible evidence, allowing vector search to explore through $w$.
This preserves a routing opportunity; reaching $v$ still depends on the retained graph and search budget.

Aggregation preserves attribute presence while discarding which record supplied each value.
For example, two contributors can supply different required labels, causing a subset check to pass even when neither record contains both labels.
Thus Marker compatibility identifies directions worth exploring, vector distance determines search priority, and exact checks on original attributes determine result eligibility.
The conservative interpretation applies to evidence represented in $S_e$, with bounded edge recovery supplementing directions rejected by Marker checks.

\subsection{Joint Search in EMA}

EMA adopts a two-stage graph search strategy.
The upper layer supports fast navigation based solely on vector distance without
attribute filtering, while the bottom layer integrates distance evaluation with attribute filtering to enable efficient \emph{joint search}.
By decoupling global navigation from predicate-aware refinement, EMA achieves both fast convergence and fine-grained filtering.

\textbf{Marker-guided joint expansion.}
As illustrated in Figure~\ref{fig:workflow}, EMA performs predicate-aware traversal in the bottom layer by leveraging Markers for early filtering.
During neighbor expansion, an edge is traversed only if its Marker is compatible with the query predicate, enabling early elimination of infeasible paths before distance evaluation.

Formally, let $\mathrm{QMarker}$ denote the Marker constructed from the query predicate $q.R$.
Recall that $m$ is the number of attributes, and we partition the attribute set into numerical and categorical subsets,
$\mathcal{A} = \mathcal{A}_{\mathrm{N}} \cup \mathcal{A}_{\mathrm{C}}$ with $\mathcal{A}_{\mathrm{N}} \cap \mathcal{A}_{\mathrm{C}} = \emptyset$.
For the $j$-th attribute, we define the attribute-level matching predicate as
\par
\begingroup
\footnotesize
\[
\mathrm{MMatch}(j) =
\begin{cases}
(\mathrm{Marker}_j \;\&\; \mathrm{QMarker}_j) \neq 0,
& \mathcal{A}^{(j)} \in \mathcal{A}_{\mathrm{N}}, \\[2pt]
(\mathrm{Marker}_j \;\&\; \mathrm{QMarker}_j) = \mathrm{QMarker}_j,
& \mathcal{A}^{(j)} \in \mathcal{A}_{\mathrm{C}},
\end{cases}
\]
\endgroup
\noindent
where $\mathrm{Marker}_j$ and $\mathrm{QMarker}_j$ denote the $j$-th attribute segment of the Marker and Query Marker, respectively.

Under conjunctive predicates, an edge is eligible for expansion if its Marker satisfies
\begin{equation}
\mathrm{MCheck(Marker)} = \bigwedge_{j=1}^{m} \mathrm{MMatch}(j) = \text{true}.
\label{eq:mmatch}
\end{equation}
Intuitively, a numerical predicate matches if at least one discretized bucket overlaps, while a categorical predicate requires full coverage of the query labels.
To accelerate Marker checking, EMA leverages SIMD instructions to perform parallel bitwise operations.

\textbf{Edge recovery under low selectivity.}
Despite Marker-guided expansion, graph connectivity may still degrade under extremely selective predicates due to the bounded out-degree $M$.
To address this issue, EMA incorporates a lightweight edge recovery mechanism.
During search, we retain edges that satisfy $\mathrm{MCheck}$ as candidates and mark mismatched edges as ineligible.
When fewer than $d_{\min}$ edges pass $\mathrm{MCheck}$, \texttt{EMA} restores the closest mismatched edges into the candidate set, ensuring a minimum out-degree of $d_{\min}$.

\texttt{EMA} jointly leverages Marker checking and bounded edge recovery to expand the reachable search region during traversal.
Markers steer exploration toward predicate-compatible directions, while edge recovery maintains sufficient local connectivity to enable traversal beyond the reach of post-filtering and two-hop expansion.


\textbf{Exact predicate verification.}
Markers provide a conservative approximation for predicate pruning and may admit false positives.
Therefore, once a node is accessed, its original attributes are used to re-evaluate the query predicate to ensure correctness.
Predicate evaluation and predicate-to-Marker translation incur negligible overhead.
Numerical predicates are evaluated via simple comparisons, while categorical attributes are represented as compact bit vectors and evaluated using SIMD-accelerated bitwise operations.
With fixed-width attribute encodings, both operations are linear in the number of predicate conditions.

\subsection{Predicate Aggregation}


In this work, we consider general Boolean predicate expressions over numerical and categorical attributes, supporting arbitrary combinations of conjunctive (\texttt{AND}) and disjunctive (\texttt{OR}) operators. This formulation subsumes conjunctive predicates, which are prevalent in real-world analytical workloads~\cite{bonifati2017analytical, neumann2011characteristic}, and aligns with the marker-based design in EMA. Specifically, the aggregation of Markers in Equation~\ref{eq:mmatch} is generalized from conjunctive evaluation to Boolean evaluation, enabling flexible predicate composition with minimal changes to the framework.

\subsection{Dynamic Support}
\label{sec:dynamic-support}
EMA maintains graph edges and their Markers jointly under insertions, deletions, and complete-record replacements.
Marker width and existing Codebook assignments remain fixed during updates.
Write $F(x)={\tt MEncode}(x.A,\mathcal{C})$.
For a node $a$, $\mathrm{level}(a)\in\{0,1\}$ denotes its highest layer.

\textbf{Insertion.}
An inserted record supplies its vector and complete attributes.
Before insertion or attribute replacement, values outside the initial numerical range use the first or last bucket without rebucketing.
Each unregistered categorical label within the configured capacity is assigned once to a bucket with the fewest assigned labels; the mapping is saved with the index and reused.
EMA finds candidate neighbors and applies Section~\ref{sec:ema-construction}: new edges include target-own encodings, and pruning aggregates attribute information into retained edges.

\textbf{Deletion.}
EMA extends IP-DiskANN's search-based local reconnection~\cite{xu2025ipdiskann} with source-specific Marker cleanup and reconstruction.
The same search discovers sources with edges to the deleted node and supplies replacement candidates, avoiding reverse-edge metadata.
Algorithm~\ref{alg:dynamic-delete} gives the single-point deletion workflow: repair planning, followed by per-source Marker cleanup and row reconstruction as needed.
Repairs complete synchronously at each layer containing the deleted node; layer~0 couples graph repair with Marker maintenance.

\begin{figure}[t]
    \centering
    \resizebox{\linewidth}{!}{\begingroup
\renewcommand{\sfdefault}{phv}
\newcommand{\dynlabel}{\sffamily\fontsize{8}{9.6}\selectfont}
\newcommand{\dynnote}{\sffamily\fontsize{7}{8.4}\selectfont}
\definecolor{dynOrange}{RGB}{242,133,39}
\definecolor{dynBlue}{RGB}{0,119,166}
\definecolor{dynGold}{RGB}{255,237,174}
\definecolor{dynBit}{RGB}{139,182,218}
\definecolor{dynRed}{RGB}{190,0,0}
\newcommand{\dynbits}[5]{%
    \begin{scope}[shift={(#1,#2)}]
        \foreach \value [count=\i from 0] in {#3,#4,#5} {
            \ifnum\value=1 \def\cellfill{dynBit}\else\def\cellfill{black!18}\fi
            \draw[fill=\cellfill,line width=0.45pt]
                (0.22*\i,0) rectangle (0.22*\i+0.22,0.28);
            \node[bit] at (0.22*\i+0.11,0.14) {\value};
        }
    \end{scope}%
}
\begin{tikzpicture}[
    x=1cm,y=1cm,
    font=\dynlabel,
    point/.style={circle,draw=black,line width=0.55pt,minimum size=4.5mm,inner sep=0pt},
    orange/.style={point,fill=dynOrange},
    blue/.style={point,fill=dynBlue},
    gold/.style={point,fill=dynGold},
    edge/.style={draw=black,line width=0.65pt,->},
    bit/.style={font=\dynnote,inner sep=0pt},
    every label/.style={font=\dynlabel,inner sep=1pt},
    cross/.style={draw=dynRed,line width=1.3pt},
    title/.style={font=\rmfamily\bfseries\fontsize{8}{9.6}\selectfont,align=center,inner sep=0pt}
]
\path[use as bounding box] (-0.20,-0.65) rectangle (7.55,3.40);
\node[font=\dynnote,anchor=east] at (1.45,3.25) {Attribute bits:};
\node[gold,minimum size=2.8mm] at (1.70,3.25) {};
\dynbits{1.95}{3.11}{1}{0}{0}
\node[orange,minimum size=2.8mm] at (3.05,3.25) {};
\dynbits{3.30}{3.11}{0}{1}{0}
\node[blue,minimum size=2.8mm] at (4.40,3.25) {};
\dynbits{4.65}{3.11}{0}{0}{1}
\foreach \offset in {0,4} {
    \begin{scope}[shift={(\offset,0)}]
        \fill[black!12]
            (0.20,0.10) ++(82:2.90)
            arc[start angle=82,end angle=15,radius=2.90]
            -- ++(195:1.088233)
            arc[start angle=15,end angle=82,radius=1.811767] -- cycle;
        \draw[dashed,line width=0.45pt]
            (0.20,0.10) ++(82:1.811767)
            arc[start angle=82,end angle=15,radius=1.811767];
        \ifnum\offset=0
            \node[point,fill=white,label=left:u] (u) at (0.20,0.10) {};
            \node[blue,label=above left:v] (owner) at (1.60,1.25) {};
            \node[gold,label=right:a] (retired) at (1.20,2.30) {};
        \else
            \draw[dashed,line width=0.55pt] (0.20,0.10) -- (1.20,2.30);
            \node[point,fill=white,label=left:u] (u) at (0.20,0.10) {};
            \node[gold,label=above left:a] (owner) at (1.60,1.25) {};
            \node[blue,label=right:v] (retired) at (1.20,2.30) {};
        \fi
        \node[orange,label=right:w] (w) at (2.45,1.35) {};
        \draw[edge] (u) -- (owner);
        \draw[cross] (1.05,2.15) -- (1.35,2.45);
        \draw[cross] (1.05,2.45) -- (1.35,2.15);
        \ifnum\offset=0
            \node[font=\dynnote,anchor=east] at (1.80,0.40) {Stored};
            \node[font=\dynnote,anchor=east] at (1.80,0.02) {Cleaned};
        \else
            \node[font=\dynnote,anchor=east] at (1.80,0.40) {Initial};
            \node[font=\dynnote,anchor=east] at (1.80,0.02) {Rebuilt};
        \fi
    \end{scope}
}
\dynbits{1.90}{0.26}{1}{1}{1}
\dynbits{1.90}{-0.12}{0}{1}{1}
\node[title] at (1.65,-0.60) {(a) Cleanup};
\begin{scope}[shift={(4,0)}]
    \node[font=\dynnote] at (0.35,1.94) {Old request};
    \dynbits{0.02}{1.53}{0}{1}{1}
    \dynbits{1.90}{0.26}{1}{0}{0}
    \dynbits{1.90}{-0.12}{1}{1}{0}
    \node[title] at (1.65,-0.60) {(b) Reconstruction};
\end{scope}
\end{tikzpicture}
\endgroup}
    \caption{Deletion examples.}
    \label{fig:dynamic-delete}
\end{figure}

\begin{algorithm}[!t]
\caption{{\tt Marker-aware Deletion}}
\label{alg:dynamic-delete}
\small
\KwIn{Two-layer graph $G$; node $a$ to delete; Codebook $\mathcal{C}$; search width $efc$; budgets $M_\ell,k_r,c$}
\KwOut{Locally updated $G$ and edge Markers}
retire $a$\;
\For{$\ell \gets 0$ \KwTo $\mathrm{level}(a)$}{
    $(Cand,Expanded)\gets{\tt Search}_\ell(a,efc)$\;
    $(P,Q)\gets{\tt Propose}_\ell(a,Cand,Expanded,k_r,c)$\;
    \ForEach{live $u\in Expanded\cup Cand\cup\operatorname{src}(P)$}{
        \If{$\ell=0$ \textbf{and} $u\in Cand$}{
            ${\tt Clean}(u,a,Cand)$\;
        }
        \lIf{$P_u=\emptyset$ \textbf{and} all $N_\ell(u)$ are live}{\textbf{continue}}
        $Nbrs\gets{\tt Live}(N_\ell(u)\cup P_u)\setminus\{u\}$\;
        $W\gets\emptyset$\;
        \If{$|Nbrs|>M_\ell$}{
            $(Nbrs,W,Q_u)\gets{\tt Prune}_\ell(u,Nbrs,Q_u,M_\ell)$\;
        }
        \If{$\ell=0$}{
            \ForEach{$v\in Nbrs$}{
                \If{$e_{(u,v)}$ is new}{
                    $e_{(u,v)}.\mathrm{Marker}\gets F(v)$\;
                }
            }
            ${\tt Recover}(u,Nbrs,W,Q_u,Cand)$\;
        }
        $N_\ell(u)\gets Nbrs$\;
    }
}
\end{algorithm}

\emph{(i) Repair planning.}
At each layer containing $a$, ${\tt Search}_\ell$ uses $a$'s old vector and construction width $efc$.
$Cand$ retains search results and distances; $Expanded$ records source IDs whose adjacency rows were examined.
${\tt Propose}_\ell$ scans their outgoing edges for $u\to a$ and ranks replacement endpoints within the original first $k_r$ positions of $Cand$.
For each discovered $u\to a$, it selects up to $c$ live targets nearest to $u$; for each live outgoing neighbor $v$ of $a$, it selects up to $c$ live sources nearest to $v$.
$P_u$ stores proposed targets for source $u$; $Q_u$ contains their Marker-transfer requests, indexed by target as $Q_{u,v}$.
Missing entries are empty.
Requests carry the replaced edge's Marker, when present, including when the proposed edge already exists.
Write $N_\ell(u)$ for $u$'s neighbor IDs and ${\tt Live}$ for liveness filtering; set expressions use node IDs.

\emph{(ii) Marker cleanup.}
Pruning transfers attribute contributions to nonincident edge Markers, so deletion requires cleanup beyond the deleted node's incident edges.
Cleanup follows Algorithm~\ref{alg:newpruning}'s source-relative RNG rule: $v$ dominates $w$ relative to $u$ when $dis(u,v)\le dis(u,w)$ and $dis(v,w)<dis(u,w)$.
For each live source $u\in Cand$, ${\tt Clean}$ selects its closest live neighbor $v$ that dominates $a$, if one exists.
On $u\to v$, it clears a Marker bit shared with $F(a)$ only if the bit is absent from $F(v)$ and no other local live witness dominated by $v$ supports it.
Witnesses range over all live candidates in $Cand\setminus\{a,u,v\}$ (Figure~\ref{fig:dynamic-delete}(a)).

\emph{(iii) Row and Marker reconstruction.}
Rows with dead-target edges or proposals are rebuilt under the layer budget $M_\ell=O(M)$.
${\tt Prune}_\ell$ sorts the pool by distance to $u$ and applies RNG-style pruning, recording direct witnesses by retained target in $W$ and adding old Markers lost through overflow pruning to $Q_u$.
Surviving old edges retain their cleaned Markers; new edges start with $F(v)$.
${\tt Recover}$ first merges live direct-witness encodings into their targets' Markers.
For a retained target $v$ missing requested bits, a live candidate $w\in Cand$ is eligible when $v$ dominates $w$ and $F(w)\land Q_{u,v}=F(w)$.
Candidates exclude the source, retained targets, and direct witnesses.
A witness already covered by a dominating retained target requires no transfer; each remaining witness contributes to its closest eligible requesting target.
Reconstruction follows current local support rather than historical pruning assignments.
Thus old Markers guide recovery but are not themselves live evidence (Figure~\ref{fig:dynamic-delete}(b)).

\textbf{Modification.}
A complete-record replacement runs as a deletion batch followed by an insertion under fresh record IDs, without slot reuse, in-place Marker patching, or a rebuild.

\section{Theoretical Analysis of EMA}


In this section, we present a theoretical analysis of EMA.
We analyze construction time and storage for EMA's two-layer graph.
We also bound the local deletion procedure, including graph repair and Marker maintenance.
We further analyze the sources of false positives introduced by Marker checking and derive theoretical bounds on their rates.
All proofs are deferred to the technical report~\cite{ema_appendix}.

\subsection{Complexity}

\begin{lemma}[Cost of MEncode and Edge Checking]
In Algorithm~\ref{alg:newpruning}, \texttt{MEncode} takes
$O(m)$ time. Testing whether $e_{(u,v)}$ previously exists takes $O(M)$ time,
as it scans the adjacency list of $u$, which contains at most $M$ neighbors.
\label{le:mencode}
\end{lemma}

\begin{lemma}[Cost of Marker-augmented pruning]
\label{le:pruning}
The time complexity of Algorithm~\ref{alg:newpruning} is $O\!\left(efc \cdot M \cdot (M+m)\right)$.
\end{lemma}

\begin{theorem}[Index construction time complexity]
\label{thm:time}
Under the parameterized graph-search model below, the total time complexity of EMA index construction is
$$O\!\left(M \cdot efc \cdot n \log n \cdot (M + m)\right).$$
\end{theorem}

The model assigns $O(efc\log n)$ examined rows to candidate search per insertion, at cost $O(M\cdot efc\log n)$.
Construction charges an additional $O(M+m)$ factor for edge checks and Marker operations (Lemma~\ref{le:pruning}).
For fixed $M$, $efc$, and $m$, the total work is $O(n\log n)$.

\begin{theorem}
The space complexity of EMA is $O(nM\cdot sm)$, where $s$ is the per-attribute Marker size and $m$ is the number of attributes.
\end{theorem}

The two graph layers store $O(nM)$ edges, with $sm$ Marker bits on each bottom-layer edge.
For fixed $s$ and $m$, total storage is $O(nM)$.

\begin{lemma}[Deletion complexity]
\label{le:deletion-cost}
Consider the local deletion procedure in Algorithm~\ref{alg:dynamic-delete} for one point $a$ in EMA's two-layer index with $n$ stored points, degree parameter $M$, search width $efc$, and $m$ attributes.
Fix the vector dimension, per-attribute encoding capacities, and reconnection budgets.
Under the same parameterized graph-search model used in the construction analysis and the standard expected constant-time model for identifier-dictionary access, its expected work is
\begin{equation}
O\!\left(M \cdot efc \cdot (M+efc\cdot m)\log n\right).
\label{eq:deletion-local-cost}
\end{equation}
\end{lemma}

The two-layer loop contributes a constant factor.
The adopted search model supplies the $O(efc\log n)$ examined-row term, with queue maintenance included in the search cost.
Each repaired row costs $O(M^2+M\cdot efc\cdot m)$ for neighbor selection and Marker reconstruction, reusing the deletion's candidate and witness sets rather than launching a fresh search.
For fixed $M$, $efc$, and $m$, the bound reduces to $O(\log n)$.
The bound applies to the logical procedure under these search and dictionary models.

\subsection{Theoretical Analysis of Marker False Positives}

A Marker false positive occurs when an edge passes early filtering although its
target violates the predicate $R$.
We analyze two source-specific mechanisms separately.

\textbf{Case (1): Dominance-induced aggregation.}
Consider the event that an edge's target fails $R$ but at least one summarized
dominated node satisfies $R$.
Let $\mu$ be the mean number of dominated nodes per edge, excluding the target;
$\mathrm{FPR}_{\mathrm{case1}}$ denotes this event's probability for a random edge.

\begin{theorem}[Case-(1) bound]
\label{thm:case1-main}
If node predicate matches are i.i.d.\ with probability $\mathrm{sel}$,
conditional on the graph, then
\[
\mathrm{FPR}_{\mathrm{case1}}
\;\le\;
(1-\mathrm{sel})\big(1-(1-\mathrm{sel})^{\mu}\big).
\]
\end{theorem}

The bound uses only the mean dominated-set size and is tight when this size is
constant across edges.
For fixed $\mu$, it vanishes as $\mathrm{sel}$ approaches either $0$ or $1$.
This event requires a matching dominated-node witness; a Marker pass assembled
from different nodes satisfying separate conjuncts is outside the bound.

\textbf{Case (2): Codebook granularity mismatch.}
For a conjunction $R=\bigwedge_{j=1}^{m} R_j$ on single-valued attributes, expand
each $R_j$ to the union of Codebook regions it intersects, and conjoin the
expanded per-attribute checks.
A partially intersected region contains both values satisfying $R_j$ and values
that do not.
The expansion cannot discard a true match, but may accept a non-matching value
that shares a region with a match.
Let $\mathrm{FPR}_{\mathrm{case2}}(R)$ be the fraction of records passing this
approximation that fail $R$.
This definition excludes neighborhood aggregation.

\begin{theorem}[Case-(2) bound]
\label{thm:case2-main}
Suppose each attribute domain is partitioned into $s$ disjoint Codebook regions,
each of probability mass at most $1/s$, and $R_j$ partially intersects at most
$b_j$ regions.
If the joint predicate selectivity is at least $\mathrm{sel}>0$, then
\[
\mathrm{FPR}_{\mathrm{case2}}(R)
\;\le\;
\frac{\sum_{j=1}^{m} b_j}
{s\,\mathrm{sel}+\sum_{j=1}^{m} b_j}.
\]
\end{theorem}

For a target rate $0<\mathrm{FP}<1$, it therefore suffices to choose
\[
s \;\ge\;
\frac{1-\mathrm{FP}}{\mathrm{FP}\,\mathrm{sel}}
\sum_{j=1}^{m} b_j.
\]
For a one-dimensional range on a contiguous partition, $b_j\le 2$.
No independence between attributes is required for this bound.

\paragraph*{Discussion}
Dominance-induced passes can preserve navigation toward nearby matching nodes,
whereas Codebook-only passes need not reflect any matching summarized node.
At fixed selectivity and boundary complexity, finer Codebooks tighten the
discretization bound, trading Marker space and checking work for finer filtering.
For a fixed error target, the sufficient resolution grows inversely with the
selectivity lower bound.
Predicates aligned with complete Codebook regions eliminate the discretization
error, but do not rule out aggregation-induced passes.
The two rates use different populations (edges versus Codebook-accepted records)
and are not additive components of the overall Marker false-positive rate.
The i.i.d.\ and bucket-mass assumptions delimit these analysis models, rather than
providing a workload-independent prediction of query throughput.

\section{Experiments}

We evaluate EMA on multi-predicate, single-attribute, negatively correlated, and complex-predicate queries, then examine ablations, dynamic updates, and resource costs.
Additional experiments, detailed evaluation protocols, and reproducibility records are available in our technical report~\cite{ema_appendix}.

\subsection{Experimental Setup}

All experiments were conducted on a high-end CPU server running Ubuntu 24.04, equipped with two Intel Xeon Platinum 8358 CPUs (64 physical cores in total), 2\,TB RAM, and HDD-backed storage for index persistence.

\textbf{Execution and metrics.}
Index construction uses one thread for \texttt{VBase} and 32 for other methods.
Queries are single-threaded except for \texttt{Milvus}, whose internal thread count is not explicitly controllable; it is therefore a system-level baseline.
We report QPS; distances are L2 or, for $\ell_2$-normalized vectors, negated inner product.

\textbf{Baselines.}
We compare with \texttt{Milvus-HNSW}~\cite{wang2021milvus}, \texttt{VBase}~\cite{zhang2023vbase}, \texttt{ACORN}~\cite{patel2024acorn}, \texttt{NaviX}~\cite{sehgal2025navix}, \texttt{Filtered DiskANN}~\cite{gollapudi2023filtered}, and \texttt{iRangeGraph}~\cite{xu2024irangegraph}.
For \texttt{ACORN} and \texttt{faiss-NaviX}, we replace eager predicate preprocessing with lazy evaluation on first access and cache the results, including predicate-checking costs in query execution.


\texttt{Filtered DiskANN} and \texttt{iRangeGraph} provide specialized label and range baselines.
The latter pre-filters on the first numerical predicate and post-filters on remaining predicates.
We exclude NHQ~\cite{wang2024efficient} because its exact-label matching does not support our subset-style predicates (query labels $\subseteq$ item attributes).

\textbf{Hyper-parameter Settings.} 
Following prior studies~\cite{peng2025dynamic, malkov2018efficient}, we set $M_{\mathrm{HNSW}}=40$ and $efc=300$ for all methods except \texttt{VBase}.
We use $\gamma=10$ for \texttt{ACORN}.
For Milvus-standalone 2.5.24 (Docker), we use numerical partition size 64; array-valued categorical attributes do not support partitioning.
\texttt{VBase} does not expose tunable hyper-parameters.

Unless stated otherwise, EMA uses a Codebook size of $s=256$, a recovery threshold of $d_{\min}=16$, and a top-layer search width of $ef_{\text{top}}=1$.

All methods have a 20-hour index construction timeout.

\textbf{Datasets.} Four datasets are used in our experiments:
\begin{table}[t]
  \caption{Dataset Statistics.}
  \label{dataset}
  \begin{tabular}{c|c|c|c|c}
    \hline
    \textbf{Dataset} & \textbf{Dim} & \textbf{Size} & \textbf{Query Size} & \textbf{Metric}\\
    \hline
    \hline
    YouTube-RGB\tablefootnote{https://research.google.com/youtube8m/download.html} & 1024 & 1M & 1000 & IP \\
    \hline
    Redcaps \tablefootnote{https://redcaps.xyz/} & 512 & 4M & 1000 & IP \\
    \hline
    SIFT \tablefootnote{http://corpus-texmex.irisa.fr/} & 128 & 10M & 1000 & L2 \\
    \hline
    Wiki\tablefootnote{https://huggingface.co/datasets/gaurav8297/navix} & 1024 & 15.4M & 1050 & IP \\
    \hline
\end{tabular}
\end{table}

\textbf{SIFT}~\cite{jegou2011searching, aumuller2020ann} contains 128-dimensional image embeddings with L2 distance.
We sample 10M vectors from SIFT-1B and assign synthetic numerical and categorical attributes for range and label queries.

\textbf{RedCaps}~\cite{desai2021redcaps} is a large-scale image--text dataset collected from Reddit, where each data point is represented by precomputed
CLIP embeddings~\cite{radford2021learning} and associated with a timestamp as a numerical attribute.

\textbf{YouTube-RGB} uses visual embeddings from the YouTube-8M dataset.
We use the upload timestamp and the number of likes as filterable attributes for range and multi-attribute queries, providing a challenging and realistic evaluation setting.

\textbf{Wiki} is adopted from \texttt{NaviX}~\cite{sehgal2025navix}, where two weakly correlated textual embedding datasets are combined.
Following NaviX's terminology, we evaluate \emph{negatively correlated queries}, where predicate-satisfying records are less likely to occur among the query's nearest neighbors than under a random selection of the same size.
This workload pairs queries about non-person entities with birth-date predicates selecting person records.
Resource records have numerical attribute zero.
We use the 50 queries provided with the dataset for this evaluation and generate an additional 1{,}000 queries via random sampling with perturbations for general experiments.

\textbf{Query workloads.}
We evaluate label+range conjunctions and mixed \texttt{AND}/\texttt{OR} predicates in the main paper.
Synthetic numerical attributes are integers sampled from $[0,100{,}000]$.
YouTube-RGB uses real timestamps and like counts; RedCaps combines timestamps with synthetic attributes.
For categorical attributes, we generate 18 labels with different probabilities and randomly assign them to all vectors.
For conjunctive queries, selectivity is evenly allocated to the two predicates to target the desired overall selectivity.
Unless noted otherwise, results target 95\% recall@10; lower-recall curves and internally multi-threaded \texttt{Milvus} results are identified separately.
In Figures~\ref{fig:rangelabelcombined} and~\ref{fig:single_attribute}, solid, dashed, and dotted curves target 95\%, 90\%, and 80\% recall@10, respectively.

\textbf{Speedup accounting.}
Speedup is EMA's QPS divided by that of the fastest displayed baseline at the same 95\% recall@10 target, including \texttt{Milvus}(MT).
Lower-recall and missing results are excluded without interpolation.
Arithmetic means equally weight unrounded pointwise ratios.

\subsection{Multi-predicate Query Performance}

This experiment tests whether EMA combines selective expansion with sufficient exploration across label+range selectivities of 1\%--100\%.
Across the label+range and composed-predicate workloads in Figure~\ref{fig:rangelabelcombined}, EMA achieves $\EMAMultiMin\times$--$\EMAMultiMax\times$ the QPS of the fastest qualifying baseline, with an arithmetic mean of $\EMAMultiMean\times$ at the same 95\% recall@10 target.

\begin{figure*}[t]
    \centering
    \input{CH5/figures/range_label_combined_qps_tikz.tex}
    \caption{QPS for multi-predicate queries.}
    \label{fig:rangelabelcombined}
    \label{fig:rangelabelhigh}
    \label{fig:rangelabel}
    \label{fig:comosed}
\end{figure*}


\textbf{High selectivity performance.}
We evaluate high-selectivity scenarios (10\%--100\%) under label+range queries.
Figure~\ref{fig:rangelabelcombined}(a) reports the results at 95\% recall.

EMA achieves the highest QPS across the evaluated datasets, with the gap generally widening as selectivity increases.
As more records satisfy the predicate, attribute checks become less selective and geometric navigation becomes increasingly important.
EMA retains RNG-style geometric pruning while using compact Marker checks to reject incompatible expansion directions before neighboring records are dereferenced.
At 100\% selectivity, predicate rejection no longer explains the advantage; graph structure and search implementation also matter.
\texttt{VBase} does not consistently reach 95\% recall; its SIFT results are reported at 90\% recall.
\texttt{Milvus} is omitted on Wiki because HDD-backed index construction does not reliably complete within the resource budget.


\textbf{Low selectivity performance.}
Figure~\ref{fig:rangelabelcombined}(b) reports QPS at 95\% recall with the selectivity levels ranging from 1\% to 10\% under label+range queries.

EMA maintains high throughput under restrictive predicates.
EMA uses evidence from pruned candidates to precheck directions before neighboring records are dereferenced, retaining non-matching intermediates whose Markers are compatible with the predicate.
When few edges pass Marker checking, bounded recovery restores the closest rejected edges until the configured local expansion threshold is met, subject to the available neighbors.
The threshold limits the number of rejected edges restored at each expansion.
\texttt{Milvus} also benefits at low selectivity from partitioning and internally parallel execution, including linear scans below 3\% selectivity.
\texttt{VBase}'s performance varies more across datasets.

\begin{figure*}[t]
    \centering
    \input{CH5/figures/single_attribute_combined_qps_tikz.tex}
    \caption{QPS for (a) range queries, (b) label queries, and (c) negatively correlated queries.}
    \label{fig:single_attribute}
    \label{fig:rangeqps}
    \label{fig:labelqps}
    \label{fig:negcor}
\end{figure*}

\subsection{Single-attribute Query Performance}
\textbf{Range filtering performance. }
Figure~\ref{fig:single_attribute}(a) compares \texttt{EMA} with \texttt{iRangeGraph} and other baseline methods under selectivities ranging from 1\% to 10\%.
\texttt{iRangeGraph} is evaluated on RedCaps and YouTube-RGB, where construction remains practical.
Across their shared 95\%-target configurations, \texttt{iRangeGraph} achieves higher QPS than EMA, while EMA outperforms \texttt{NaviX}.
The gap between \texttt{iRangeGraph} and EMA narrows as selectivity increases; construction and storage costs are reported in Table~\ref{tab:index_time}.
Its range-specific sub-indexes allow pre-filtering to focus search on the requested interval.
EMA instead searches a shared graph using discretized summaries, trading this specialization for support of mixed predicates.
As the interval broadens, pre-filtering excludes fewer records, reducing its search-space advantage.

\textbf{Label filtering performance.} Figure~\ref{fig:single_attribute}(b) reports low-selectivity label filtering.
\texttt{Filtered DiskANN} on YouTube-RGB and \texttt{ACORN} on RedCaps are reported at 80\% recall and are excluded from equal-recall comparisons.
EMA achieves higher QPS than the qualifying 95\%-target baselines across the displayed selectivities.
For sparse matching labels, EMA's Markers retain label evidence beyond the immediate target, allowing non-matching records to remain routing intermediates.
Bounded recovery supplements this guidance when few outgoing edges pass the checks.

\subsection{Negatively Correlated Queries}
\label{sec:negative-experiments}

This workload tests routing when geometric proximity and predicate satisfaction are misaligned.
For the Wiki queries described in the setup, nearby resource records fail the birth-date predicates selecting person records.
Similarity-directed exploration can therefore spend substantial work in non-matching neighborhoods; local expansion must reach beyond these neighborhoods to obtain matching candidates.

Figure~\ref{fig:negcor}(c) shows that EMA reaches 95\% recall across the evaluated selectivities.
Under the evaluated settings, \texttt{ACORN} remains below 60\% recall and \texttt{NaviX} plateaus around 90\%; \texttt{VBase} also fails to reach 95\%.
The plotted \texttt{NaviX} and \texttt{VBase} curves use 90\% and 80\% recall, respectively.
EMA obtains 9.79--68.57 QPS as selectivity increases from 1\% to 23\%.

EMA uses evidence from dominated candidates to expand non-matching routing intermediates.
When too few edges pass Marker checking, bounded recovery provides additional expansion opportunities beyond the non-matching neighborhoods.

\subsection{Complex Predicates}

To test composition of numerical and categorical constraints within one index, we use
\par
\nopagebreak[4]
\begingroup
\footnotesize
\setlength{\abovedisplayskip}{3pt}
\setlength{\abovedisplayshortskip}{3pt}
\[
p = \bigl( \texttt{num} \in [a_1, b_1] \wedge \texttt{cate} \supseteq L_1 \bigr)
\lor \bigl( \texttt{num} \in [a_2, b_2] \wedge \texttt{cate} \supseteq L_2 \bigr).
\]
\endgroup
\noindent
Each branch conjoins a range with a label-subset condition; the two branches are joined by \texttt{OR}.
We vary the overall selectivity from 1\% to 10\% at 95\% recall@10.
Selectivity is measured over the complete disjunction, including overlap between its branches.

Figure~\ref{fig:rangelabelcombined}(c) shows EMA's throughput advantage for this composed-predicate workload.
EMA evaluates the branches using the same per-attribute Codebooks and edge Markers, combining their bit-level checks before dereferencing neighboring records.
This reuses the filtering guidance without building a separate graph for each predicate combination.

\subsection{Ablation of Core Mechanisms}
\label{sec:ablation}

\begin{figure*}[t]
    \centering
    \begin{minipage}[t]{0.49\textwidth}
    \vspace{0pt}
    \centering
    \input{CH5/figures/ablation_tikz.tex}\par\nointerlineskip
    \captionof{figure}{YouTube-RGB ablation at nominal selectivities and 95\% recall@10.}
    \label{fig:ablation}
    \end{minipage}\hfill
    \begin{minipage}[t]{0.49\textwidth}
    \vspace{0pt}
    \centering
    \input{CH5/figures/parameter_sensitivity_tikz.tex}\par\nointerlineskip
    \captionof{figure}{RedCaps4M parameter sensitivity at 95\% recall@10.}
    \label{fig:parameter-sensitivity}
    \end{minipage}
\end{figure*}

\begin{figure*}[t]
    \noindent
    \begin{minipage}[t]{0.60\textwidth}
    \vspace{0pt}
    \footnotesize
    \setlength{\tabcolsep}{4pt}
    \captionof{table}{Index size (GB) and time (s).}
    \label{tab:index_time}
    \begin{tabular}{@{}c*{8}{|c}@{}}
    \hline
    \multirow{3}{*}{\textbf{Algorithm}}
    & \multicolumn{2}{c|}{\textbf{YoutubeRGB1M}}
    & \multicolumn{2}{c|}{\textbf{Redcaps4M}}
    & \multicolumn{2}{c|}{\textbf{SIFT10M}}
    & \multicolumn{2}{c}{\textbf{Wiki15.4M}} \\
    & \multicolumn{2}{c|}{\textbf{(3.9\,GB)}}
    & \multicolumn{2}{c|}{\textbf{(7.7\,GB)}}
    & \multicolumn{2}{c|}{\textbf{(4.9\,GB)}}
    & \multicolumn{2}{c}{\textbf{(59\,GB)}} \\
    \cline{2-9}
    \rule{0pt}{\dimexpr\ht\strutbox+0.5pt\relax}
    & Size & Time & Size & Time & Size & Time & Size & Time \\
    \hline
    \hline
    \texttt{VBase} & - & 1,429 & - & 4,162 & - & 7,549 & - & 48,609 \\
    \hline
    \texttt{Milvus} & 40 & 221 & 40 & 368 & 40 & 565 & - & - \\
    \hline
    \texttt{ACORN} & 4.4 & 973 & 9.7 & 2,412 & 9.9 & 4,110 & 67 & 21,085 \\
    \hline
    \texttt{NaviX} & 4.2 & 51 & 8.9 & 73 & 7.9 & 82 & 64 & 391 \\
    \hline
    \texttt{Filtered DiskANN} & 4.0 & 474 & 8.4 & 2,219 & 18 & 4,794 & 62 & 7,070 \\
    \hline
    \texttt{iRangeGraph} & 5.4 & 4,582 & 13.9 & 11,001 & 24.9 & 22,002 & - & - \\
    \hline
    \texttt{EMA} & 6.6 & 293 & 19 & 591 & 32 & 1,137 & 101 & 3,396 \\
    \hline
    \end{tabular}
    \end{minipage}\hfill
    \begin{minipage}[t]{0.37\textwidth}
    \vspace{0pt}
    \centering
    \input{CH5/figures/ema_dynamic_compact_tikz.tex}\par\nointerlineskip
    \captionof{figure}{SIFT10M updates (95\% recall).}
    \label{fig:ema_updates}
    \end{minipage}
\end{figure*}

\newcommand{\EMAAblationFullLowMin}{1.37}
\newcommand{\EMAAblationFullLowMax}{1.55}
\newcommand{\EMAAblationMarkerForty}{2.34}
\newcommand{\EMAAblationMarkerSixty}{1.60}
\newcommand{\EMAAblationMarkerEighty}{1.16}
\newcommand{\EMAAblationDistanceReduction}{45.67}
\newcommand{\EMAAblationPredicateReduction}{81.51}
\newcommand{\EMAAblationExpansionIncrease}{20.96}

Figure~\ref{fig:ablation} uses mixed predicates, 128-bit Markers, and $d_{\min}=16$.
\texttt{EMA} combines Marker-guided expansion and bounded edge recovery; \texttt{No Recovery} disables recovery.
\texttt{Post-filtering} is the same-implementation post-filtering control.
Its QPS comes from the component study; the other throughput curves use the fixed-graph run.
QPS is bracket-interpolated; speedups are seven-round paired medians.
\texttt{EMA} improves throughput at 1\%--10\%.
The strict in-filtering control admits directly reached bottom-layer neighbors for distance evaluation and queue insertion after exact predicate checks pass.
On the shared graph, \texttt{No Recovery} achieves \EMAAblationMarkerForty$\times$ strict in-filtering's QPS at 40\%.
At 1\%--10\%, both encounter incomplete query batches (NC); recovery enables \texttt{EMA} to reach the target.
Panel~(b) normalizes work to post-filtering on the same saved graph as \texttt{EMA}.
At 1\%, \texttt{EMA} reduces distance evaluations and predicate checks by \EMAAblationDistanceReduction\% and \EMAAblationPredicateReduction\%, despite \EMAAblationExpansionIncrease\% more node expansions.
Filtering benefits vanish at 100\%.

\subsection{Parameter Sensitivity}
\label{sec:parameter-sensitivity}

\textbf{Marker width.}
Wider Markers distinguish finer attribute regions but enlarge the summary carried by each edge.
The QPS peak at 128 bits rather than 256 in Figure~\ref{fig:parameter-sensitivity}(a) shows that encoding resolution alone does not determine query efficiency: the value of more selective prechecks must be weighed against summary access and checking work.

\textbf{Recovery threshold.}
Recovery re-admits geometrically close edges rejected by Marker checks, trading selective expansion for additional routes.
Without recovery, search misses the recall target at 1\% selectivity, whereas the largest tested budget is slower than moderate thresholds throughout Figure~\ref{fig:parameter-sensitivity}(b).
Recovery must therefore supply alternatives when the eligible neighborhood is sparse while limiting the additional candidate work introduced by bypassing Marker checks.

\subsection{Dynamic Maintenance}

\textbf{Workloads and stages.}
Figure~\ref{fig:ema_updates} evaluates insertion, synchronous deletion, and complete-point update on 128-dimensional SIFT10M with its original numerical and categorical attributes and squared L2 distance.
Stage 0 is the baseline; each of stages 1--5 applies 1M logical record changes.
Insertion grows a 5M-record index to 10M; deletion starts from that final checkpoint and shrinks it back to 5M in random batches.
Complete-point update starts from the original 5M checkpoint and replaces 1M original records per stage, changing both vectors and attributes under fresh record IDs without slot reuse.
Its live size remains 5M, and by stage 5 all initially indexed records have been replaced.
Graph and Marker repair finish synchronously before querying.

\textbf{Evaluation protocol.}
All traces reuse 1,000 query vectors and fixed mixed predicates, with mean selectivity 9.79\% on the full corpus; selectivity may vary with the live set.
To track query performance over successive record changes, all three traces use 128 Marker bits per attribute and a routing minimum degree of 8, held fixed from their initial checkpoints onward.
Codebooks remain fixed, maintenance uses 32 threads, and queries use one physical core with $k=10$.

\textbf{Query throughput.}
Figure~\ref{fig:ema_updates}(b) evaluates each checkpoint at 95\% external-ID recall@10 against exact filtered GT over its current live set.
We calibrate $ef$ and interpolate median QPS only within a measured recall bracket.
Insertion and deletion change the live corpus size, so their opposing QPS trends do not isolate maintenance-induced degradation.
Fixed-size replacement instead tests whether routing and filtering guidance remain effective as records turn over.
Deletion can invalidate both edges and evidence on nonincident Markers; EMA combines local reconnection with Marker cleanup and live-witness reconstruction (Section~\ref{sec:dynamic-support}).
After complete replacement, EMA retains \EMAUpdateRetention\% of its initial query throughput without rebuilding the graph.

\textbf{Index modification cost.}
Figure~\ref{fig:ema_updates}(a) reports wall time per 1M-record batch, including synchronous maintenance and in-call preparation and cleanup, but excluding initial construction, evaluation, and checkpoint I/O.
Deletion requires reconnection and live-witness checks before clearing shared Marker bits, consistent with its higher cost than insertion.
Replacement combines deletion and insertion rather than an in-place overwrite; its nearly flat batch times show limited cost growth over this trace despite accumulating retired IDs.

\subsection{Index Construction and Space}
Table~\ref{tab:index_time} reports index construction time and index size. Results for \texttt{Milvus} and \texttt{iRangeGraph} on Wiki are omitted as index construction does not reliably complete within the time or resource budget at this scale.
EMA requires more construction time and storage than \texttt{NaviX}: on SIFT10M, 1,137 versus 82 seconds and 32 versus 7.9\,GB; on Wiki, 3,396 versus 391 seconds and 101 versus 64\,GB.
\texttt{Milvus}'s disk footprint is estimated with its official sizing tool\footnote{\url{https://milvus.io/tools/sizing}}; actual allocation follows fixed segment granularity.

Marker encoding and propagation add construction work, and per-edge summaries occupy index space.
These summaries retain pruning-derived evidence for repeated query-time checks across numerical, categorical, and composed predicates.
The shared representation trades preprocessing and storage for selective traversal across these workloads.
Predicate-specific indexes offer a different balance for applications centered on one filter type, as illustrated by the range-only results.



\section{Related Work}

\textbf{Filtering ANNS Algorithms.}
Early work on Filtering ANNS can be traced back to RII~\cite{matsui2018reconfigurable}, which considers subset search and can be viewed as an early form of pre-filtering.

For numerical range filtering, SeRF~\cite{zuo2024serf} stores edges for query ranges during HNSW construction, and ESF~\cite{peng2025dynamic} further supports dynamic insertions.
DIGRA~\cite{jiang2025digra} combines a dynamic multi-way tree with ANN indexes to support range filtering and updates.
RangePQ~\cite{zhang2025rangepq} integrates range filtering with product-quantization-based indexes and supports dynamic updates.
Other numerical range methods focus on organizing ANN sub-indexes.
$\beta$-WST~\cite{engels2024approximate} organizes sub-indexes in a window search tree.
UNIFY~\cite{liang2024unify} merges multiple subgraphs using cross-subgraph edges.
iRangeGraph~\cite{xu2024irangegraph} dynamically selects subgraphs from a binary tree for range-filtered search.

For categorical label filtering, AIRSHIP~\cite{zhao2022constrained} considers label distributions and constructs a label-aware HNSW index.
Filtered DiskANN~\cite{gollapudi2023filtered} incorporates label coverage into graph pruning.
NHQ~\cite{wang2024efficient} combines vector and attribute distances for graph-based hybrid queries with equality constraints.

Predicate-agnostic Filtering ANNS methods accommodate different predicates during traversal.
ACORN~\cite{patel2024acorn} includes two-hop neighbor expansion to navigate under restrictive predicates.
NaviX~\cite{sehgal2025navix} adapts first- and second-hop exploration using global and local selectivity.

These approaches differ in how they obtain guidance for graph expansion.
Filtered DiskANN uses attribute-aware pruning to determine which edges survive, whereas ACORN and NaviX inspect additional neighbors during traversal.
EMA reuses attribute evidence from geometrically pruned candidates by associating it with retained edges.
These edge-level Markers precheck expansion directions before target records are dereferenced, retaining directions with predicate-compatible evidence from pruned candidates even when the immediate targets fail the predicate.

\textbf{System-Level Filtering ANNS Methods.}
Relational databases with vector support provide an integration point for vector search and relational predicates.
PostgreSQL-based systems such as PGVector~\cite{pgvector}, PASE~\cite{yang2020pase}, and VBase~\cite{zhang2023vbase} combine vector access with predicate evaluation.
PASE performs iterative search to improve reliability under filtering constraints, while VBase relaxes monotonicity to support filtered queries.
Vector and graph databases also integrate filtering into their search engines.
AnalyticDB-V~\cite{wei2020analyticdb} supports both pre-filtering and post-filtering based on cost estimation.
Milvus~\cite{wang2021milvus} partitions indexes into multiple sub-indexes to support subset search and parallel execution.
Weaviate~\cite{weaviate2024} integrates ACORN~\cite{patel2024acorn} to enable efficient predicate filtering.
NaviX is implemented as a native vector index in Kuzu~\cite{feng2023kuzu,sehgal2025navix}.

\textbf{Dynamic Index Maintenance.}
IP-DiskANN~\cite{xu2025ipdiskann} supports in-place graph updates through search-based local reconnection without a reverse-edge index.
EMA follows this local-reconnection approach and adds source-specific Marker cleanup and reconstruction.
In EMA, a deleted record may have contributed attribute bits to nonincident Markers during earlier pruning.
Local repairs use target encodings and surviving local witnesses to clean affected Marker bits and reconstruct summaries where needed.
This couples topology repair with maintenance of the filtering evidence distributed across retained edges.

\section{Conclusion}

EMA is an edge-centric Filtering ANNS framework supporting Boolean combinations of numerical range and categorical label predicates.
It transfers attribute evidence from geometrically pruned candidates to compact edge Markers, guiding selective expansion through useful non-matching intermediates.
Our analysis characterizes construction and maintenance costs and bounds source-specific Marker false positives under explicit assumptions.
Synchronous local graph and Marker maintenance supports insertion, deletion, and complete-record replacement.
On the evaluated multi-predicate workloads, EMA achieves $\EMAMultiMin\times$--$\EMAMultiMax\times$ the QPS of the fastest baseline at the 95\% recall@10 target.

\clearpage
\bibliographystyle{IEEEtran}
\bibliography{sample}

\end{document}